\documentstyle[12pt]{article}

\def\NPB#1#2#3{Nucl. Phys. B{#1} (19#2) #3}
\def\PLB#1#2#3{Phys. Lett. B{#1} (19#2) #3}

\def\ov{\overline}
\def\s2{\frac{1}{\sqrt2}}

\def\oh{\frac{1}{2}}
\def\beq{\begin{equation}}
\def\eeq{\end{equation}}
\def\beqa{\begin{eqnarray}}
\def\eeqa{\end{eqnarray}}
\def\IP{\relax{\rm I\kern-.18em P}}
\def\IC{\relax\hbox{\kern.25em$\inbar\kern-.3em{\rm C}$}}

\def\IZ{\relax{\rm Z\kern-.4em  Z}}
\def\cp#1{\relax\ifmmode {\IP\kern-2pt{}_{#1}}\else
$\IP\kern-2pt{}_{#1}$\fi}
\def\stu{\hbox{$S$-$T$-$U$}}

\def\re{\hbox{Re\,}}

\topmargin
-2cm
\textwidth
15cm
\textheight
24cm
\oddsidemargin
1cm
\evensidemargin
1cm
\begin{document}
\date{}
%\vskip 1cm
\title{ Heterotic/Heterotic Duality in $D=6,4$}

\author{G. Aldazabal \thanks{Permanent Institutions:
CNEA, Centro At\'omico
Bariloche, 8400 S.C. de Bariloche, and CONICET, Argentina.} $^1$,
A. Font \thanks{On sabbatical leave from
Departamento de F\'{\i}sica, Facultad de Ciencias,
Universidad Central de Venezuela. Work supported in part by the
N.S.F. grant PHY9511632 and the Robert A. Welch Foundation.} $^2$,
L.E. Ib\'a\~nez$^{1}$ and F. Quevedo$^3$ \\[4mm]
$^1$ \normalsize Departamento de F\'{\i}sica Te\'orica, \\[-3mm]
\normalsize Universidad Aut\'onoma de Madrid, \\[-3mm]
\normalsize Cantoblanco, 28049 Madrid, Spain.
\\[3mm]
 $^2$\normalsize Theory Group, Department of Physics,\\[-3mm]
\normalsize The University of Texas,\\[-3mm]
\normalsize Austin, TX 78712, USA.
\\[3mm]
 $^3$ \normalsize Theory Division, CERN, 1211 Geneva 23,
Switzerland.
}
\maketitle
\vspace{-4.6in}
%\hspace{1in}
\begin{flushright}
CERN-TH/96-35, FTUAM-96/05 \\[-2mm]
UTTG-02-96 \\[-2mm]
hep-th/9602097 \\
\end{flushright}
\vspace{3.5in}
\begin{abstract}

We consider $E_8\times E_8$ heterotic compactifications on  $K3$  and
$K3\times T^2$. The idea of heterotic/heterotic duality in $D=6$
has difficulties for generic compactifications since for large
dilaton values some gauge groups acquire negative kinetic terms.
Recently Duff, Minasian and Witten (DMW) suggested a solution to this
problem which only works
if  the compactification is performed assuming the presence of
symmetric gauge embeddings on both $E_8$'s . We consider an
alternative in which
asymmetric  embeddings are possible  and the wrong sign of
kinetic terms for large dilaton value
is a signal of  spontaneous symmetry breaking.
Upon  further toroidal compactification to $D=4$, we find that
the duals in the DMW case correspond to  $N=2$ models
in which the $\beta $-function of the
different group factors  verify  ${\beta }_\alpha=12$
whereas the asymmetric solutions that we propose have
${\beta }_\alpha=24$.
We check the consistency of these dualities
by studying the different large $T,S$ limits of the
gauge kinetic function.
Dual $N=1$, $D=4$  models can also be obtained by the operation of
appropriate freely acting twists, as shown in specific examples.

\end{abstract}
%\vskip .1in

\begin{flushleft}
CERN-TH/96-35 \\
February 1996 \\
\end{flushleft}

\maketitle

%\end{document}
\newpage

\section{ Introduction}

In a recent paper \cite{dmw},
Duff, Minasian and Witten (DMW) have provided evidence
for the existence of a heterotic/heterotic duality in six dimensions.
This duality relates a weakly coupled $E_8\times E_8$ theory
compactified on $K3$ to a strongly coupled $E_8\times E_8$ theory
obtained upon compactification on a differently realized $K3$.
Heterotic/heterotic duality was first conjectured in
refs.~\cite{Dufflu, Duffkhuri, Minasian, Duff} motivated by
heterotic/fivebrane duality \cite{strom} in $D=10$. It may also be
understood
\cite{dmw}
as two alternative (dual)
ways of looking at the compactification of the  $D=11$ $M$-theory  on
$K3\times S^1/Z_2$.

In the present note we reexamine the idea of $D=6$ and $D=4$
heterotic/heterotic duality concerning different aspects.
First, we would like to propose that this duality  is not
only present in  $E_8\times E_8$ compactifications in which
equal instanton numbers are  embedded in both factors
but is also present  in certain asymmetric cases. In these
new cases the gauge group is also Higgsed away at generic points in
hypermultiplet moduli space but, unlike the DMW case,
the dual gauge particles do not  have their origin in small instanton
effects.
Secondly, we examine the $D=4$ theories obtained in both cases
upon further toroidal compactification. We find that the resulting
\nobreak $N=2$ theories have $\beta_{\alpha } ^{N=2}=12$ in the DMW
case and
$\beta_{\alpha} ^{N=2}=24$ in the new proposed duals for {\it all }
perturbative gauge groups appearing at enhanced symmetry points.
We  show how  the $N=2$ models obtained in the case
of non-symmetric embeddings  are related  to  certain chains of
heterotic/type II  duals studied in ref.~\cite{afiq}.
We also   study the large $T$ and/or $S$ limit of the $N=2$ gauge
kinetic
functions for those theories and obtain consistency with duality
only precisely for the above mentioned  values of the $\beta
$-functions.
Finally, we also show how $D=4$, $N=1$ duality can be derived
upon twisting by appropriate freely-acting symmetries of the
$N=2$ models.

\section{$D=6$ heterotic/heterotic duality}

One of the most intuitive hints \cite{Minasian} for the existence
of a $D=6$  heterotic/heterotic duality is the way in which  the
anomaly
eight-form $I_8$ factorizes into the product of four-forms in $D=6$.
In fact, $I_8=X_4{\tilde X}_4$, with
\beqa
\label{uno}
X_4\ & =& \frac{1}{4(2\pi )^2} \left( trR^2\ -\ v_{\alpha
}trF_{\alpha
}^2\right) \\
{\tilde X}_4\ & =& \frac{1}{4(2\pi )^2} \left ( trR^2\ -\ {\tilde
v}_{\alpha }trF_{\alpha }^2 \right) \quad ,\nonumber
\label{lasx}
\eeqa
where ${\alpha} $ runs over the gauge groups in the model.
This very symmetric form of $I_8$  suggests a duality under which
one exchanges the tree-level Chern-Simons contribution to the
Bianchi identity $dH={\alpha }'(2\pi )^2X_4$ with the one-loop
Green-Schwarz corrections to the field equations
$d{\tilde H} ={\alpha }'(2\pi )^2{\tilde X}_4$. In these expressions
$v_{\alpha }$  is a (positive) tree-level coefficient which is
essentially the
Kac-Moody  level. On the other hand the coefficients ${\tilde
v}_{\alpha }$
are  associated to the Green-Schwarz mechanism, they depend on the
massless spectrum of the model and they can be positive, negative or
zero.
 The problem with a naive duality
exchanging both terms is that  a generic $D=6$ model  yields
values for  ${\tilde v}_{\alpha }$ that have the wrong sign
(see the formulae for ${\tilde v}_{\alpha }$ below).

In fact, independently of any duality hypothesis, the existence of
negative values for ${\tilde  v}_{\alpha }$ is in general problematic
since,
as shown by Sagnotti \cite{Sagnotti}, the exact dilaton dependence of
the gauge kinetic terms in $D=6$
is given (in the Einstein frame) by the expression:
\beq
L_{gauge}^{D=6}\ =\
-\frac{(2\pi )^3}{8{\alpha }' } {\sqrt {G}}
\left (v_{\alpha }e^{-\phi /2} \ +\ {\tilde v}_{\alpha }e^{\phi /2}
\right)
trF_{\alpha }^2
\label{sagno}
\eeq
where $\phi $ is the six-dimensional dilaton field.
This expression shows that there are  only  tree-level and one-loop
contributions and that the corresponding coefficients are given by
$v_{\alpha }$, ${\tilde v}_{\alpha }$.  It is obvious that, if   any
of the
${\tilde v}_{\alpha }$'s is negative (as  happens in the generic
case),
for some value of the dilaton the theory becomes inconsistent
and hence one cannot continuously extrapolate towards strong
coupling.

Since the coefficients $v_{\alpha }$ and ${\tilde v}_{\alpha }$ play
an
important role in the discussion below, we will first briefly review
the relevant formulae that allow their computation in the $D=6$,
$N=1$
theory.  At the Kac-Moody level $k_\alpha =1$,
the coefficient $v_{\alpha }$ is given by
$v_\alpha = 2,1,\frac{1}{3},\frac{1}{6},
\frac{1}{30}$ for $\alpha = SU(N), SO(2N), E_6, E_7, E_8$,
respectively \cite{Erler,dmw}.
The derivation of the $\tilde v_\alpha$ involves
constraints on the massless spectrum that follow from the
conditions of anomaly cancellation \cite{GSW, Erler}.
These conditions include cancellation of the term $tr\, R^4$ that
requires $(n_h - n_v)=244$, where $n_h$ ($n_v$) is the number of
hyper(vector)multiplets.

For groups with an independent fourth-order
Casimir, cancellation of $tr\, F^4$ imposes the constraint
\beq
T_{\alpha }\ =\ \Sigma _i \ n_i t^i_{\alpha} \quad ,
\label{anom}
\eeq
where $n_i$ refers to the number of hypermultiplets in representation
$R_i$. $T_{\alpha }$
and $t^i_{\alpha }$ are group-theoretical quantities that appear
when decomposing $TrF_{\alpha }^4$ into factorized and non-factorized
terms.
In the notation of ref.~\cite{Erler},
\beq
TrF_{\alpha }^4=T_{\alpha} trF_{\alpha }^4 + U_{\alpha
}\left(trF_{\alpha}^2\right)^2
\quad\quad ; \quad\quad
tr_{R_i}F_{\alpha }^4 = t^i_{\alpha} trF_{\alpha }^4 + u^i_{\alpha
}\left(trF_{\alpha}^2\right)^2.
\label{norm}
\eeq
For example, for $SU(N)$ with $N\geq 4$,
$T_{SU(N)}=2N$  and $t^i_{SU(N)}= 1$, $(N-8)$, $\oh(N^2-17N+54)$,
$\frac1{6}(N-4)(N^2-23N+96)$ for the
fundamental and $2, 3, 4$-index antisymmetric representations
respectively.

In ref.~\cite{Erler}, an analysis of the total anomaly led to
general expressions for the coefficients $\tilde v_\alpha$ at the
level $k_\alpha = 1$. We will be mostly interested in the
following
\beqa
{\tilde v}_{SU(N)} & = & n_{a2}+(N-4)n_{a3} +
\oh (N-4)(N-5)n_{a4}  -2
\quad\quad (N\geq 4) \nonumber \\
{\tilde v}_{SO(2N)} & = & 2^{(N-6)} n_s -2
\quad\quad   (N\geq 3) \nonumber \\
{\tilde v}_{SU(2)} & = &  \frac{n_2-16}{6} \quad\quad ; \quad\quad
{\tilde v}_{SU(3)}  =  \frac{n_3-18}{6}  \nonumber \\
{\tilde v}_{E_6} & = & \frac{n_{27}-6}{6} \quad\quad ; \quad\quad
{\tilde v}_{E_7}  =  \frac{n_{56}-4}{6} \quad\quad ; \quad\quad
{\tilde v}_{E_8}  = - \frac{1}{5} \quad . %\nonumber
\label{vtild}
\eeqa
Here $n_{aj}$ refers to the number of $j$-index antisymmetric $SU(N)$
tensors, $n_s$ to the number of $SO(2N)$ spinorials and the rest of
the notation is self-explanatory. Notice that almost always $n_i$
also
counts the complex $\ov{R}_i$ representation.
The representations appearing in ~(\ref{vtild}) are those allowed by
unitarity at $k_\alpha = 1$ and potentially massless.
The number of fundamentals, $n_f$, has been
eliminated in the first two groups by virtue of eq.~(\ref{anom}).

Observing formulae (\ref{vtild}) one immediately realizes the
many possibilities for negative
${\tilde v}_{\alpha }$ that would in turn lead to negative kinetic
terms (for large $\phi $) in eq.(\ref{sagno}). To start with,
whenever there is a gauge group without charged hypermultiplets,
${\tilde v}_{\alpha}$ is negative.
Consider for example a  $D=6$ compactification obtained embedding
$SU(2)$ bundles with instanton numbers $s$ and $s'$ respectively in
each $E_8$. Modular invariance dictates that $s+s'=24$. The
standard embeding choice ($s=24$, $s'=0$) yields a model
with gauge group $E_7\times E_8$, and hypermultiplets transforming
like $10(56;1)+65(1;1)$.  In this case ${\tilde v}_{E_7}=1$ and
${\tilde v}_{E_8}=-1/5$.  More generally, a sufficient amount of
matter
hypermultiplets is required for $\tilde v_\alpha > 0$. In the case at
hand, for $s,s' \not= 0$, the gauge group is $E_7\times E_7'$,
and  $s+s'=24$ implies $n_{56} + n_{56'} =8$. Since we need
$n_{56}\geq 4$ and $n_{56'}\geq 4$ for non-negative ${\tilde
v_\alpha}$'s
we are forced to have $n_{56}=n_{56'}=4$.
This corresponds to $s=s'=12$ and ${\tilde v}_{\alpha }=0$
for both  $E_7$'s, {\it i.e.},  the symmetric embedding.
It is trivial to check that any non-Abelian subgroup obtained by
Higgsing with hypermultiplets also has ${\tilde v}_{\alpha }=0$.

The solution proposed by DMW to the wrong sign kinetic term
problem is  precisely
to restrict to compactifications with ${\tilde v}_{\alpha}=0$.
Since this possibility is obviously not symmetric
under the exchange of $v_{\alpha }$ and ${\tilde v}_{\alpha}$,
it requires  that the dual gauge group be generated by
non-perturbative (small instanton) effects as suggested in
\cite{smallins} .
This hypothesis is consistent with the fact that the known examples
of gauge groups generated by these non-perturbative effects
verify $v_{\alpha }=0$ (unlike the perturbative ones, which
obviously  have $v_{\alpha}>0$). The proposal can be
justified \cite{dmw}
by considering this duality as arising from two (dual) ways of
looking
at the compactification of the $D=11$ $M$-theory on $K3\times
S^1/Z_2$.
The two dual $D=6$ theories correspond to $E_8\times E_8$
heterotic compactifications on $K3$ with symmetric gauge bundles in
both $E_8$'s. However, the original and the dual $K3$ are not
identical,
the dual corresponding to a  $K3$ orbifold of type $T^4/ Z_2$.

A generic compactification of the $E_8\times E_8$ heterotic string on
$K3$, with equal $SU(2)$ instanton numbers on both $E_8$'s, gives
rise
to an $N=1$, $D=6$ model with gauge group $E_7\times E_7$ and
hypermultiplets transforming as $4(56,1)+4(1,56)$  $+62(1,1)$.
For generic vev's of the hypermultiplets the gauge symmetry is
completely broken and one is left with $244$ hypermultiplets
and no vector multiplets.
Thus there is only some gauge group
at enhancing points in the moduli space. The idea is that those
points in moduli
space are different from the ones at which the non-perturbative
gauge group may be generated, so that  one does not have to deal
simultaneously with perturbative and non-perturbative gauge groups,
which are supposed to be dual.

Since the dual $K3$ has the structure of
(some sort of) $T^4/Z_2$ orbifold \cite{dmw},
it is interesting to see whether one can construct a $Z_2$
orbifold with the appropriate characteristics. Let us then consider
the
twist $\theta $ acting on the complexified $T^4$ coordinates
$z_{1,2}$ as
$\theta (z_1,z_2) =(-z_1,-z_2)$  and let us look for an embedding
on $E_8\times E_8$ symmetric in both $E_8$'s.
It is easy to check that on the $E_8\times E_8$ lattice
there is no modular invariant order-two shift $A$
that is symmetric in both $E_8$'s. Nonetheless,
a $Z_2$ orbifold with symmetric spectrum can be constructed as
follows.
To begin with, consider just the standard embedding of $\theta$ by
the
gauge lattice shift
\beq
A = \left(\oh,\oh,0,0,0,0,0,0\right)\times \left(0,
 \cdots,0\right) \quad .
\label{shiftA}
\eeq
Although this does not look very symmetric, the model is symmetrized
if, in addition, we turn on the Wilson line
\beq
a_1 = \left(\oh,\oh,0,\cdots,0\right)\times
\left(\oh,\oh,0,\cdots,0\right) \quad ,
\label{wline}
\eeq
along, say, the first compact dimension
\footnote
{For  simple rules to find the massless spectrum in
orbifolds with underlying quantized Wilson lines see
ref.~\cite{sierra}.}.
The resulting model then
has gauge group $(E_7\times SU(2))^2$, four singlet hypermultiplets
in the
untwisted sector and hypermultiplets transforming as
$4(56,1;1,1)+4(1,1;56,1)$ $+16(1,2;1,1)+16(1,1;1,2)$  in the twisted
sector. The  model is perfectly symmetric and thus
constitutes a good candidate for the searched dual $T^4/Z_2$
orbifold.
Notice that upon Higgsing of the two $SU(2)$'s, the particle content
of
this model corresponds to the particle content of a $K3$
compactification
with equal  $SU(2)$ instanton
numbers embedded in both $E_8$'s, as described above.

We would now like to consider  an alternative to the above
explicit realization of heterotic/heterotic duality that
could occur even if the gauge bundles are embedded differently
in both $E_8$'s.  Indeed,  a negative ${\tilde v}_{\alpha }$ in
eq.(\ref{sagno})
is not by itself problematic.  It is only problematic if one insists
in
going to large $\phi $.  A way to avoid this dangerous limit would be
to
remain at finite $\phi $ values (so that the theory is well defined)
and
Higgs away the gauge groups with negative ${\tilde v}_{\alpha}$
by giving vev's to the  hypermultiplets in the theory. Once all the
dangerous gauge groups have been Higgsed away, one can safely take
the  large $\phi $ limit  and go to strong coupling.
This point of view gives a physical  interpretation to the large
$\phi $ limit  of theories with some negative ${\tilde v}_{\alpha }$,
namely, for large $\phi $ there is a phase transition in which the
corresponding gauge group is spontaneously broken.

It is obvious that this mechanism is not always functional. In
particular, there must be
sufficient hypermultiplets  in the theory to Higgs away completely
the wrong sign gauge groups. Consider for example a $E_7\times E_7'$
theory. Since $E_7$ has $133$ generators one needs a
minimum of 3($56$)'s to Higgs completely one $E_7$.
We are thus led to the unique choice of an embedding with
$s=14$ and $s'=10$ that gives rise to hypermultiplets transforming as
$5(56,1)+3(1,56')$$+62(1,1)$. From (\ref{vtild}) we find
${\tilde v}_{E_7}= 1/6$ and ${\tilde v}_{E_7'}= -1/6$.
However, $E_7'$ can be completely Higgsed away with the $3(1,56')$,
leaving behind a $D=6$ heterotic model
with gauge group $E_7$ and hypermultiplets transforming
as $5(56)+97(1)$ in which
\beq
{\tilde v}_{E_7 }\ =\   v_{E_7}\ =\ \frac{1}{6} \quad .
\label{nos}
\eeq
Notice that not only is ${\tilde v}$ positive but it is
also equal to $v$. Thus in this model explicit  heterotic/heterotic
duality along the original ideas of
refs.~\cite{Dufflu, Duffkhuri, Minasian, Duff}
seems possible. It is easy to check that
any non-Abelian group obtained from this $E_7$ by Higgsing
still verifies  the condition $v_{\alpha }={\tilde v}_{\alpha }$.
Notice that  for generic points in hypermultiplet moduli space
the gauge group is again spontaneously broken and, as in the case of
symmetric embedding,  244 hypermultiplets remain massless.

This is not the only example in which Higgsing the groups with
$\tilde v_\alpha < 0$ is feasible.
We will now describe a different type of $D=6$ heterotic
compactification in which
this procedure works as in the previous example.  It is a $D=6$
model based on the standard $Z_6$ orbifold  with a
$E_8\times E_8$ embedding given by
$V= \frac16(1,1,1,1,-4,0,0,0)\times \frac16(1,1,1,1,1,-5,0,0)$.
The resulting gauge group is $SU(5) \times SU(4) \times U(1)
\times
SU(6) \times SU(3) \times SU(2)$ and the massless hypermultiplets are
\beqa
\label{spz6}
\theta^0&:&(1,\ov{4},-5;1,1,1)+(10,4,1;1,1,1)+(1,1,0;6,3,2)\
+2\, (1,1,0;1,1,1)\nonumber\\
\theta^1&:& (1,1,\frac{10}3;1,3,2)+(1,4,-\frac53;6,1,1)
 + 2\, (1,1,\frac{10}3;6,1,1)  \nonumber\\
\theta^2&:& 5\, (1,\ov{4},\frac53;1,\ov{3},1)+
4\, (\ov{5},1,-\frac43;1,\ov{3},1) \nonumber\\
\theta^3&:& 3\, (1,6,0;1,1,2) + 5\, (5,1,-2;1,1,2) \quad ,
\eeqa
where ${\theta }^n$ indicates to which twisted sector
the hypermultiplet belongs. Using (\ref{vtild}) we find
${\tilde v}_{SU(5)}=2$, ${\tilde v}_{SU(4)}=4$,
${\tilde v}_{SU(6)}=-2$, ${\tilde v}_{SU(3)}=6$
and ${\tilde v}_{SU(2)}=8$.
Only $SU(6)$ has wrong sign but,
since for $SU(N)$ one has $v_{SU(N)}=2$, only $SU(5)$
looks promising to produce a self-dual theory.
One can check that indeed there are appropriate hypermultiplets to
Higgs completely all gauge groups but $SU(5)$. In this way we obtain
a $D=6$ $SU(5)$ model with hypermultiplets transforming
as $4(10)+22(5)+118(1)$ and ${\tilde v}_{SU(5)}=v_{SU(5)}=2$.
As in the previous example, for generic points in hypermultiplet
moduli
space there is a $D=6$ heterotic theory with no vector multiplets
and $244$ hypermultiplets.

It is worth remarking that in this type of $D=6$ models in
which the unbroken gauge group (at enhanced points) has
 ${\tilde v}_{\alpha }=v_{\alpha }$, one expects the occurrence of
heterotic/heterotic duality as a consequence of
$D=10$ string/fivebrane duality in the same spirit as originally
formulated in refs.~\cite{Dufflu, Duffkhuri, Minasian, Duff}.
In particular, no small instanton effects seem to be required to
obtain
duality.

\section{$D=4$  heterotic/heterotic duality}

Let us now consider the  $D=4$ heterotic models obtained upon further
compactification of the above $D=6$  heterotic duals  on a 2-torus.
This case was considered in ref.~\cite{Duff} and briefly mentioned in
ref.~\cite{dmw}. The resulting
$N=2$ theory has the usual toroidal vector multiplets
$S,T,U$ related to the coupling constant and the size and shape of
the 2-torus. When the $D=6$ theory is
dimensionally reduced to four dimensions,
the underlying duality exchanges the roles of
$S$ and $T$ \cite{Duff, Pierre}. Including mirror symmetry on the
torus,
one thus expects complete \stu\ symmetry in this type of vacua
\cite{Duff, dmw, Rahmfeld, klm, Duffrev, cclmr}.
Thus, on top of the usual perturbative $SL(2,\IZ)_T\times
SL(2,\IZ)_U$
dualities, a non-perturbative $SL(2,\IZ)_S$  $S$-duality \cite{filq}
is expected. This $N=2$ model has the toroidal $U(1)^4$ as
generic gauge group, and as matter, $244$ neutral
hypermultiplets (it corresponds to the heterotic construction of
model $B$
of ref.~\cite{Kachru}). At particular points in moduli space,
enhanced gauge groups such as $E_7\times E_7$ can appear.

A natural question is the following: What is the $D=4$ equivalent of
the ${\tilde v}_{\alpha }=0$  or  ${\tilde v}_{\alpha }=v_{\alpha }$
conditions we had in $D=6$  in order to
have heterotic/heterotic duality? It turns out that the equivalent
condition
in $D=4$ can be phrased as a condition on the $N=2$ $\beta
$-functions
of the gauge groups present at enhanced points in moduli space.
Indeed, the $N=2$, $D=4$ $\beta $-function of a given
gauge factor can be written in terms of
the corresponding $D=6$ ${\tilde v}_{\alpha }$ coefficient.

As an exercise let us consider the case of $SU(N)$ ($N\geq 4$)
with the representations that appear at level $k=1$.
With the notation introduced before, the $N=2$ $\beta $-function is
given by
\beq
{\beta }_{SU(N)}^{N=2}\ =\ -2N + n_f  +n_{a2}(N-2) +
\oh n_{a3}(N-3)(N-2) + \frac{1}{6}n_{a4}(N-4)(N-3)(N-2).
\label{bet}
\eeq
Now, the $D=6$ anomaly factorization condition in eq.~(\ref{anom} )
implies
\beq
2N\ =\ n_f + n_{a2}(N-8)+ \oh n_{a3}(N^2-17N+54)
+ \frac{1}{6} n_{a4}(N-4)(N^2 - 23N + 96).
\label{anomN}
\eeq
Combining these two expressions with that  for ${\tilde
v}_{\alpha }$
in eq.~( \ref{vtild}) we arrive at
\beq
{\beta }_{SU(N)}^{N=2}\ =\ 12\ +\ 6\, {\tilde  v}_{SU(N)} \quad .
\label{betasem}
\eeq
For the other groups in eq.~(\ref{vtild}) we obtain a similar result.
More precisely,
\beq
\beta_\alpha^{N=2} \ = \ 12\left(1 +  \frac{\tilde
v_\alpha}{v_\alpha}\right) \quad .
\label{betagen}
\eeq
Thus, the condition to get heterotic/heterotic duality in  $N=2$,
$D=4$ models reads
\beqa
\beta_\alpha^{N=2} &=&  12 \quad\quad  \hbox{(symmetric  $E_8\times
E_8$
embeddings)} \nonumber \\
\beta_\alpha^{N=2} &=&  24 \quad\quad \hbox{(non-symmetric
$E_8\times E_8$ embeddings)} \quad .
\label{dos}
\eeqa
Notice that in both cases the $N=2$ models are
non-asymptotically free.
Notice also that the $\beta$-functions of all groups
must be equal. In the first case ($\beta_{\alpha}$ =\, 12),
consistently with the DMW hypothesis in $D=6$, there
should be points in moduli space in which new gauge groups
of a non-perturbative origin should appear.
Those are required to obtain full duality.  In the second case
($\beta_{\alpha}$ =\, 24) this is not expected but explicit
duality should be apparent.

One can think of the following  consistency check of the proposed
ideas. We know the form of the holomorphic
$N=2$ gauge kinetic function $f_{\alpha }$ for the gauge groups
inherited from
$E_8\times E_8$.  For a $K3\times T^2$ compactification of the type
discussed here one has \cite{Kaplouis, deWit}
\beq
f_{\alpha }\ =\ k_{\alpha }S_{inv}\ - \
\frac{{\beta }^{N=2}_{\alpha }}{4\pi } \log \left(\eta (T)\eta
(U)\right)^4 \quad ,
\label{efe}
\eeq
where $\eta $ is the Dedekind function and $S_{inv} $ is given by:
\beq
S_{inv}\equiv S-\frac{1}{2}\partial_T\partial_U
h^{(1)}(T,U)-\frac{1}{2\pi}\log \left(J(T)-J(U)\right)
+{\rm const}.
\label{sinv}
\eeq
Here $h^{(1)}(T,U)$  is the moduli-dependent one-loop correction to
the $N=2$ prepotential ${\cal F}$. More explicitly,
\beq
{\cal F} \ = \ -STU+h^{(1)}(T,U) + \cdots \quad ,
\label{prep}
\eeq
where the ellipsis stands for terms that depend on matter fields, and
for simplicity  we do not include other moduli fields besides
$S,T,U$.
Now, we know that the large-$T$ limit of $f_\alpha$ must reproduce
the result in eq.~(\ref{sagno}).
It is not clear that this
follows from eqs.~(\ref{efe}) and (\ref{sinv}).
We want to show that indeed this is the case
if and only if $\beta_\alpha^{N=2} =12\left(1 +  \frac{\tilde
v_\alpha}{v_\alpha}\right)$.
To this purpose, we need to
know the asymptotic large-$T$ limit of $S_{inv}$ and, hence, of
$h^{(1)}$.

The one-loop correction $h^{(1)}(T,U)$ was explicitly computed
in ref.~\cite{HM} for the $K3$ representation in terms of
a $Z_2$ orbifold with standard embedding. The authors find
\beq
h^{(1)}(T,U)=-\frac{1}{(2\pi)^4}{\cal L}(T,U)-
\frac{c(0)\zeta(3)}{32\pi^4}-\frac{U^3}{12\pi}
\qquad \re T> \re U \quad ,
\label{ache}
\eeq
where
\beq
{\cal L}(T,U)\equiv L_{i_3}\left ( e^{-2\pi  (T-U)}\right)
+\sum_{k,l}c(kl)\, L_{i_3}\left ( e^{-2\pi  (kT+lU)}\right) \quad .
\label{ele}
\eeq
Here $L_{i_3}(x)\equiv \sum_1^\infty x^j/j^3$.
For the second Weyl chamber ($\re U>\re T$) the expression
for $h^{(1)}(T,U)$ is similar, exchanging $T\leftrightarrow U$.
Both expressions are defined up to quadratic terms that
have no physical significance.
Here the coefficients $c(n)$ are defined by the expansion
of the modular form $F(M)\equiv E_6 E_4/\eta^{24}=\sum_{n=-1}^\infty
c(n)\, q^n $ with $q\equiv e^{-2\pi M}, M=U,T$.
 In particular $c(0)$ coincides with the difference
between the
number of vector and hypermultiplets ($n_v-n_h=4-244=-240$ in these
models).

Since our $E_7\times E_7$ model of the previous section is just the
same
$Z_2$ model with standard embedding considered in \cite{HM},
with the addition of a discrete Wilson line, the moduli dependent
part
of the prepotential will remain the same. Therefore
equations (\ref{prep}), (\ref{efe}) and (\ref{ache}) define the full
perturbative prepotential of our model. From
 eqs.~({\ref{ache}) and ({\ref{ele}) one can show that
$\partial_T\partial_U h^{(1)}(T,U)\rightarrow 0$ for large $T$.
Since $\log J(T)\rightarrow 2\pi T$ one then finds for $k_\alpha=1$
\beq
\lim_{T\rightarrow \infty } f_{\alpha } \ =\
S\  +T \left(\frac{\beta_{\alpha }^{N=2}}{12}\ -\ 1\right)\ =\
S\ +\  \frac{{\tilde v}_{\alpha }}{v_{\alpha }} \ T \quad ,
\label{limites}
\eeq
which is just the $D=4$ version of formula (\ref{sagno}).
We thus see that if any of the $\beta_\alpha^{N=2}$ is smaller
than  12, the large-$T$ limit
gives rise to gauge kinetic terms of the wrong sign.
Actually, before Higgsing, the relation with $\tilde v_\alpha$
implies that if there is any group with $\beta_\alpha^{N=2}\, > 12$,
there must be another group with $\beta _\alpha^{N=2}\, < 12$,
leading inescapably to negative kinetic terms.
More generally, $\beta _\alpha^{N=2}\,< 12$ is
the $D=4$ equivalent of the wrong sign problem signalled
in $D=6$ by $\tilde v_\alpha \, < 0$.

Notice the different large-$T$ behavior of the
two heterotic/heterotic dualities under consideration.
In the one  proposed  by DMW  one has ${\tilde v}_{\alpha }=0$
and $f\rightarrow S$. In the alternative, based on an non-symmetric
embedding, one has $f\rightarrow S+T$, a $S\leftrightarrow T$
invariant result.
We should remark that in this last case, the calculation of \cite{HM}
does not apply, but the limit in eq.(\ref{limites}) still holds
\cite{deWit}.

Heterotic/heterotic duality tells us that there should be
a heterotic dual model with the roles of
$S$ and $T$ exchanged.
In particular, the dual kinetic function ${\tilde f}_{\alpha }$
is obtained by making the replacement
$T\leftrightarrow S$ in  eq.(\ref{efe}) . Thus
\beq
{\tilde f}_{\alpha }\ =\ k_{\alpha }T_{inv}\ - \
\frac{{\beta }^{N=2}_{\alpha }}{4\pi } \log (\eta (S)\eta (U))^4
\quad .
\label{efedual}
\eeq
We find for $k_\alpha=1$
\beq
\lim_{S\rightarrow \infty } {\tilde f}_{\alpha } \ =\
T\  +S \left(\frac{\beta_{\alpha }^{N=2}}{12}\ -\ 1\right)\ =\
T\ +\  \frac{{\tilde v}_{\alpha }}{v_{\alpha }} \ S \quad ,
\eeq
which again shows the special roles of $\beta
_{\alpha}^{N=2}\,=12,24$.
We can have non-perturbative gauge groups only for
$\beta _{\alpha }^{N=2}\,=12$  whereas only for
$\beta _{\alpha }^{N=2}\,=24$ can we have the standard
perturbative limit.

There is an interesting connection between  the class of models
with ${\beta }^{N=2}_{\alpha }=24, 12 $ discussed here and
$D=4$ heterotic/type II
duality. Indeed, the models discussed in this section are just
a particular class of $E_8\times E_8$ compactifications
on $K3\times T^2$  with appropriate  gauge bundles.
The particularity of both types of heterotic/heterotic duals
is that the gauge symmetry coming from $E_8\times E_8$
can be Higgsed away completely.
As remarked in  ref.~{\cite{afiq}, any heterotic model  of this kind
necessarily  yields $244$ hypermultiplets  due to anomaly
constraints.
In addition there are four $U(1)$'s coming from
three vector multiplets (which contain $S$,$T$ and $U$) plus the
graviphoton.
This particle content was denoted as $(244,4)$ in
ref.~\cite{Kachru}.

In ref.~\cite{Kachru}, a candidate type II
dual for the theory with  $(244,4)$ particle
content was proposed as a compactification
on a Calabi-Yau hypersurface in $\IP(1,1,2,8,12)$,
which is a $K3$ fibration.  The heterotic dual was constructed
as a $K3\times T^2$ compactification with symmetric embedding, as
discussed in section one.
On the other hand, we proposed  a different heterotic
construction for this dual in ref.~{\cite{afiq}. This was based  on
the $Z_6$ compactification described in the previous section.
The interest of such a construction is that we were able to identify
a chain of heterotic/type II duals continuously connected by Higgsing
to the $(244,4)$ model. The same chain can be derived from the
$E_7\times E_7$ model with instanton numbers $s=14$ and $s'=10$.
Sequential Higgsing \cite{afiq} produces models with particle
content $(139,7)$, $(162,6)$, $(191,5)$ and finally  $(244,4)$.
The remarkable point is that for all these models,
that have $v_\alpha = \tilde v_\alpha$ at points of enhanced gauge
symmetry, we were able to identify {\it type II} dual candidates
as compactifications on certain $K3$ fibrations \cite{klm}.
Thus, at least for the last four elements of the chain,
there are three dual descriptions, namely,
a type II compactification and two dual heterotic descriptions.
It would be certainly interesting to study more deeply the different
connections among the various dualities appearing in the
above models.

\section{$N=1, D=4$  heterotic/heterotic duality}

To obtain $N=1$, $D=4$ heterotic models, one can perform
a $K3\times T^2$ compactifications along with some twist that
preserves just one supersymmetry.  If the twist acts freely on
$K3\times T^2$ it is reasonable to expect that the \stu\
permutation invariance symmetry should remain  and
that  the strongly coupled limit of the resulting $N=1$ model
should be given  by the weakly coupled limit of an
analogous $N=1$ model with the roles of $S$ and $T$ exchanged. Since
we want a free action, the natural choice is to mod out by a
$Z_2$ twist $\omega $ that acts on $K3$ as the (freely-acting)
Enriques
involution and on $T^2$ by a reflection $T^2\rightarrow -T^2$. In
this way
there will be one unbroken supersymmetry. Modular invariance requires
that the twist should be accompanied by some action on the gauge
group.

Although this is a general procedure that should in principle work
for any $K3\times T^2$, let us apply it to our explicit symmetric
construction of section 1, in which we consider $T^4/Z_2$ instead of
$K3$.
In this case, the action of $\theta$, with embedding by the shift $A$
in (\ref{shiftA}) and the Wilson line in (\ref{wline}), is
supplemented by the action of $\omega$ with some gauge embedding.
This action need not be symmetric in both $E_8$s as long as the
twist $\omega$ is freely-acting.

In practice we then consider an $N=1$, $Z_2\times Z_2$ orbifold
with action on the three complex coordinates of $T^6$ given by
\beqa
\theta \left(z_1, z_2,z_3\right) & = & \left(-z_1,-z_2,z_3\right)
\nonumber \\
\omega \left(z_1,z_2,z_3\right) & = & \left(z_1,-z_2, -z_3\right)\ +\
\left(\oh,\oh,0\right) \quad .
\label{z2z2}
\eeqa
The twist $\theta$ gives the original $K3\times T^2$ whereas
$\omega $ corresponds to the simultaneous action
of the Enriques involution and a reflection of $z_3$ \cite{hls}.
Again, modular invariance requires that $\omega$ come
along with some action in the gauge degrees of freedom. The  shift
\beq
B=\left(0,\oh,\oh,0,0,0,0,0\right)\times
\left(0,\oh,\oh,\oh,\oh,0,0,0\right)
\label{shiftB}
\eeq
can be shown to verify all modular invariance constraints.

With the above ingredients we then obtain an $N=1$
model with gauge group $E_6\times U(1)^2\times SU(8)\times U(1)$.
The massless spectrum includes the untwisted moduli $T_3\equiv T$
and $U_3\equiv U$. There are no charged multiplets in the untwisted
subsectors
associated to the first two complex planes. In the third-plane
untwisted
subsector we find  $(27+{\overline {27}}; 1)$ $+(1;70)+4(1;1)$ under
$E_6\times SU(8)$. In the $\theta $-sector we find
$4(27+{\overline {27}};1)$$+4(1; 28+{\overline {28}})$$+72(1;1)$.
There are no massless particles in the
sectors twisted by $\omega $ and $\theta \omega $,  since
$\omega $ acts freely.

The specific $N=1$ example outlined above was derived
from an $N=2$ theory with symmetric embedding. However,
the method employed in its construction can in principle be
applied to obtain $N=1$ examples from $N=2$ models with
non-symmetric embedding and hypermultiplet content allowing
for complete Higgsing.

As stated before, when $\omega$ acts freely, the resulting $N=1$
model is expected to have a strongly coupled limit
whose dynamics can be described by a similar
but weakly coupled $N=1$ model in which $S$ and $T$ are exchanged.
As in the $N=2$ case, a natural simple check of this idea can be
performed
by comparing the gauge kinetic functions of the original and the dual
theory. At first sight this does not look very promising, since
the unbroken gauge groups in the $N=1$ models have in general
different $N=1$ $\beta $-functions ($b_{E_6}^{N=1}=-6$
and $b_{SU(8)}^{N=1}=10$ in the above orbifold example).
But let us proceed with the argument, this objection notwithstanding.

In $N=1$ models the  gauge kinetic function
$f^{N=1}_{\alpha }$ is also equal to $k_\alpha S$ at tree level
and has only one-loop corrections. However, as shown in
ref.~\cite{Kaplouis},
the one-loop term only receives (moduli-dependent) contributions
from
sectors of the orbifold with extended $N=2$ supersymmetry. Moreover,
the
contribution of these $N=2$ subsectors is precisely of the form
given in eq.(\ref{efe}), with the coefficient of the log term
proportional to the $N=2$ $\beta $-function of the $N=2$ subsectors
of the
theory. Since $\omega $ acts freely, the only  $N=2$
subsector is the  underlying initial $N=2$, $T^4/Z_2\times T^2$
orbifold.
Thus, when $\omega $ is freely-acting we have the result
\beq
f^{N=1}_{\alpha } \ =\  f^{N=2}_{\alpha }\ \ ;\ \
 {\tilde f}^{N=1}_{\alpha } \ =\  {\tilde f}^{N=2}_{\alpha } \quad .
\label{efeuno}
\eeq
Hence, the $N=1$ gauge kinetic function is consistent with duality
and  \stu\ permutation symmetry.
Non-perturbative information about the K\"ahler potential
of this class  of models should in principle be extractable
by an appropriate truncation of the results for the
prepotential of the underlying $N=2$ theory.

Several comments concerning this class of $N=1$ models are in order.
First, notice that, although the underlying $N=2$ parent model
was asymptotically non-free, the $N=1$ gauge groups can be
asymptotically free and have unequal $\beta $-functions.
This fact, could allow us to perform a truly
non-perturbative analysis of the gaugino condensation process
in $N=1$ models and its possible relation to the
breakdown of supersymmetry.

Secondly, this class of $N=1$ models is a
relatively constrained class.  Indeed, the underlying
$N=2$ model is essentially uniquely determined as the
model with 4 vector multiplets and 244 hypermultiplets.
Of course, this model, which just has $U(1)^4$
gauge invariance at generic points in moduli space,
can get enhanced gauge symmetries as large
as $(E_7\times SU(2))^2$ at some points.
Nevertheless, the $N=1$ class of models obtained by twisting
is not so much constrained since there is a variety of
freely acting twists that can be effected
(in particular different gauge embeddings are possible).
Thirdly, note  that these models, like their
$D=6$ parent, have no unbroken gauge group
left for generic values of the chiral field scalars.
They look quite similar to the $N=1$ examples built
in refs.~\cite{hls, Vafawitten}.

There is another potentially interesting way to derive $N=1$
heterotic/heterotic dual pairs. As we said, in $D=6$ we
have a duality relating $E_8\times E_8$ heterotic compactifications
on different realizations of $K3$, say $K3$ and $K3'$,
with symmetric gauge embeddings.
In analogy with a similar situation in heterotic/type II
duality \cite{stst}, it is reasonable to conjecture that if we
compactify  the $E_8\times E_8$ heterotic
down to $D=4$ on a Calabi-Yau manifold
that happens to be {\it a $K3$-fibration over} $\cp1$
\cite{klm, Vafawitten}, the result should be dual
to a compactification on another
Calabi-Yau manifold corresponding to {\it a $K3'$-fibration over}
$\cp1$.
It would be quite interesting to find explicit examples of this kind
of conjectured dual pairs.

\vskip0.6cm
\newpage

\centerline{\bf Acknowledgements}
\bigskip

We acknowledge useful conversations with
G.L. Cardoso,  V. Kaplunovsky, A. Klemm, P. Mayr,
 R. Minasian and  A. Uranga.
G.A. thanks the Departamento de F\'{\i}sica Te\'orica
at UAM for hospitality, and the Ministry of Education
and Science of Spain
as well as CONICET (Argentina) for financial support.
A.F. thanks CONICIT (Venezuela) for a research grant S1-2700
and CDCH-UCV for a sabbatical fellowship.
L.E.I. thanks CICYT (Spain) for financial support.

%\newpage

%\section{References}

%\newpage

\end{document}